\documentclass[10pt,aps,prl,floatfix,superscriptaddress,twocolumn,showpacs,amsmath,amssymb,reprint]{revtex4-2}
\usepackage[T1]{fontenc}
\usepackage[latin9]{inputenc}
\usepackage{verbatim}
\usepackage{textcomp}
\usepackage{amsmath}
\usepackage{graphicx}

\makeatletter

\newcommand{\lyxmathsym}[1]{\ifmmode\begingroup\def\b@ld{bold}
  \text{\ifx\math@version\b@ld\bfseries\fi#1}\endgroup\else#1\fi}

\PassOptionsToPackage{caption=false}{subfig} 
\usepackage{hyperref}
\hypersetup{
breaklinks=true,
colorlinks=true,
citecolor=blue,
linkcolor=blue,
filecolor=blue,
urlcolor=blue
}
\usepackage{xcolor}
\usepackage{ulem}
\IfFileExists{lmodern.sty}{\usepackage{lmodern}}{}
\def\@fnsymbol#1{\ensuremath{\ifcase#1\or *\or *\or \dagger\or \ddagger\or
   \mathsection\or \mathparagraph\or \|\or **\or \dagger\dagger
   \or \ddagger\ddagger \else\@ctrerr\fi}}

\makeatother

\newcommand{\authnote}{VND and JBL contributed equally to this work. \\ 
Correspondence may be addressed to \href{mailto:vduarte@pppl.gov}{vduarte@pppl.gov} and \href{mailto:lestzj@fusion.gat.com}{lestzj@fusion.gat.com}.}
\usepackage{xspace}
\newcommand{\Alfven}{Alfv\'{e}n\xspace}
\newcommand{\Alfvenic}{Alfv\'{e}nic\xspace}

\usepackage{babel}
\begin{document}
\title{Shifting and splitting of resonance lines due to dynamical friction
in plasmas}
\author{V. N. Duarte}
\thanks{\authnote}
%\email{vduarte@pppl.gov}

\address{Princeton Plasma Physics Laboratory, Princeton University, Princeton,
NJ, 08543, USA}
\author{J. B. Lestz}
\thanks{\authnote}
%\email{lestzj@fusion.gat.com}

\address{General Atomics, San Diego, CA, 92121, USA}
\address{Department of Physics and Astronomy, University of California, Irvine,
CA, 92697, USA}
\author{N. N. Gorelenkov}
\address{Princeton Plasma Physics Laboratory, Princeton University, Princeton,
NJ, 08543, USA}
\author{R. B. White}
\address{Princeton Plasma Physics Laboratory, Princeton University, Princeton,
NJ, 08543, USA}
\date{\today}
\begin{abstract}

A quasilinear plasma transport theory that incorporates Fokker-Planck
dynamical friction (drag) and pitch angle scattering is self-consistently
derived from first principles for an isolated, marginally-unstable
mode resonating with an energetic minority species. It is found that
drag fundamentally changes the structure of the wave-particle resonance,
breaking its symmetry and leading to the shifting and splitting of
resonance lines. In contrast, scattering broadens the resonance in
a symmetric fashion. Comparison with fully nonlinear simulations shows
that the proposed quasilinear system preserves the exact instability
saturation amplitude and the corresponding particle redistribution
of the fully nonlinear theory. Even in situations in which drag leads
to a relatively small resonance shift, it still underpins major changes
in the redistribution of resonant particles. This novel influence
of drag is equally important in plasmas and gravitational systems.
In fusion plasmas, the effects are especially pronounced for fast-ion-driven
instabilities in tokamaks with low aspect ratio or negative triangularity,
as evidenced by past observations. The same theory directly maps to
the resonant dynamics of the rotating galactic bar and massive bodies
in its orbit, providing new techniques for analyzing galactic dynamics.

\end{abstract}
\maketitle

\textit{Introduction}.\textemdash Resonances are channels for non-adiabatic
energy exchange between particles and waves in kinetic systems. In
linear theory, resonant interactions occur when a specific synchronization
condition is satisfied exactly. In reality, however, this condition
is smeared out due to effects such as finite mode amplitude, turbulence,
and collisions. In plasmas, the effect of resonance broadening has
been historically investigated in strong turbulence \citep{Dupree1966,Weinstock1968,Salat1988,Ishihara1992,krommes_2015}
and quasilinear \citep{Berk1995LBQ,Callen2014PoPCoulombCollsions,Bian2014Broadening,duarte2019collisional,catto_2020,catto_2021}
approaches, with important implications to the dynamics of plasma
echoes \citep{SuOberman1968} and \Alfven waves \citep{WhiteDuarteGorMengPoP2019}.
Previously unexplored aspects, but similarly important for influencing
the resonant particle relaxation, are mechanisms responsible for the
shifting and splitting of discrete resonances.

The competition between convective (dynamical friction, also known
as drag) and diffusive (scattering) collisions plays a key role in
determining the dynamical behavior of wave-particle resonant systems
\citep{Lilley2009PRL}. For instance, it is crucial to explaining
the observation of \Alfven eigenmode frequency chirping in tokamaks
\citep{DuarteAxivPRL}. Although global, non-resonant effects of drag
are well known \citep{Gaffey1976JPP}, the influence of drag on the
structure of narrow resonant layers has not previously been determined.

In this work, we explore kinetic instabilities close to their threshold
to analytically show that coherent Fokker-Planck drag breaks the symmetry
of the resonances with respect to their original location. This occurs
not only by shifting and splitting them but also by altering the relative
strength of regions in the vicinity of a resonance. The skewed dependence
of the resonance due to drag is in turn responsible for significant
modifications to the particle distribution function. 
Moreover, it is demonstrated that the fully nonlinear collisional
kinetic system naturally reduces to the form of a quasilinear (QL)
theory in the limit of marginal stability and when stochastic processes
dominate the relaxation.

An important application of these findings in fusion plasmas is the
forecasting of deleterious fast ion transport by \Alfven eigenmodes.
Fully nonlinear simulations are numerically costly and therefore it
is of practical interest to develop reduced models, such as quasilinear
theory, capable of reproducing essential features of more complete
descriptions \citep{GorelenkovDuarteRBQNF2018,PodestaPPCF2014Kick}.
The drag-induced modifications of the instability saturation level,
resonance function, and fast ion redistribution are anticipated to
be more significant in tokamaks with low aspect ratio or negative
triangularity, as evidenced by past observations of a greater propensity
for dynamical \Alfven Eigenmode behavior in these configurations \citep{Gryaznevich2006NF,Podesta2012NF,Petrov2015JPP,McClementsFredrickson2017,Van_Zeeland_2019},
consistent with theoretical predictions \citep{Lilley2010,DuartePoP2017,DuarteNF2018Likelihood}.

A deep connection exists between kinetic processes in plasmas and
those present in self-gravitating systems, as both are well-described
by mean field theories governed by long-range, inverse square laws.
In fact, the respective distributions obey the same evolution equation
in phase space, giving rise to analogous phenomena such as Landau
damping and resonant relaxation in both plasmas \citep{Landau1946,ONiel1965,Mazitov1965}
and self-gravitating systems \citep{Lynden-Bell1962,Sweet1962,Mikhailovskii1972,Ryutov1999}.
Consequently, understanding the structure of collisional wave-particle
resonances is equally relevant to kinetic plasma instabilities and
galactic dynamics, for instance, in determining the torque applied
to the rotating galactic bar by orbiting heavy bodies \citep{Hamilton2022}.

\textit{Self-consistent transport theory with dynamical friction and scattering}.\textemdash{}
Krook or scattering collisions are known to lead to an anti-symmetric
modification of the particle distribution around a resonance \citep{duarte2019collisional}.
As will be demonstrated, drag introduces a distinct, asymmetric response. Hence, we investigate the 1D nonlinear dynamics of an electrostatic
wave resonating with a hot minority species, \textit{i.e.} the canonical
bump-on-tail problem, in the presence of both scattering and drag.
The generalization to more complicated geometries and to waves of
any polarization can be achieved with the methods of Refs. \citep{BerkPPR1997,duarte2019collisional}.
The wave field is represented as $E(x,t)=\textup{Re}\left(\hat{E}\left(t\right)e^{i\left(kx-\omega t\right)}\right)$,
with complex amplitude $\hat{E}(t)$. The hot minority species is
described via a distribution function $f(x,v,t)$ with an initial
discrete resonance at $kv_{\text{{res}}}=\omega$. For resonances
narrow with respect to the velocity scale of the distribution, it
is sufficient to evaluate the Fokker-Planck coefficients at the resonance
center and the kinetic equation becomes \citep{BerkPPR1997,Lilley2009PRL}
\begin{multline}
\frac{\partial f}{\partial t}+v\frac{\partial f}{\partial x}+\frac{1}{k}\textup{Re}\left(\omega_{b}^{2}e^{i\left(kx-\omega t\right)}\right)\frac{\partial f}{\partial v}=C[f-F_{0}],\label{eq:KineticEq}
\end{multline}

where 

\begin{equation}
C[g]=\frac{\nu^{3}}{k^{2}}\frac{\partial^{2}g}{\partial v^{2}}+\frac{\alpha^{2}}{k}\frac{\partial g}{\partial v}.
\end{equation}
Here $F_{0}(v)$ is the equilibrium distribution function, which is assumed to have an approximately constant positive slope
in the vicinity of the resonance. The rate of mode drive at $t=0$ in the absence of damping is proportional
to this slope: $\gamma_{L,0}=\left(2\pi^{2}q^{2}\omega/mk^{2}\right)\partial F_{0}/\partial v$,
while the mode damping rate due to interaction with the background
thermal plasma is given by a constant $\gamma_{d}.$ The nonlinear
bounce frequency for deeply-trapped resonant particles, $\omega_{b}\left(t\right)\equiv\sqrt{qk\hat{E}\left(t\right)/m}$,
is a convenient measure of the mode amplitude.

The coefficients $\nu$ and $\alpha$ are the effective scattering
and drag collision frequencies, which are enhanced with respect to
the $90\lyxmathsym{\textdegree}$ pitch angle scattering rate and
the inverse slowing down time \citep{SuOberman1968,BerkBreizmanPRL1992,Callen2014PoPCoulombCollsions,DuartePoP2017}.
(see \footnote{See Supplemental Material {[}url{]}, Sec. 1, which includes \citep{SuOberman1968,BerkBreizmanPRL1992,Callen2014PoPCoulombCollsions,DuartePoP2017}}
for heuristic arguments). Due to the narrowness of the resonances,
the resonant particle dynamics can be strongly controlled by collisions
even if their non-resonant dynamics hardly feel their effect.

Due to spatial periodicity, the distribution can be assumed of the
form $f\left(x,v,t\right)=f_{0}\left(v,t\right)+\underset{l=1}{\overset{\infty}{\sum}}\left(f_{l}\left(v,t\right)e^{il\left(kx-\omega t\right)}+c.c.\right)$.
In the collisional regime, an asymptotic expansion exists in orders
of the small parameter $\left|\omega_{b}^{2}\right|/\nu^{2}\ll1$.
Re-writing Eq. \eqref{eq:KineticEq} in orders of such a small parameter
naturally implies the ordering $\left|f_{0}'^{(0)}\right|\gg\left|f_{1}'^{(1)}\right|\gg\left|f_{0}'^{(2)}\right|,\left|f_{2}'^{(2)}\right|$
\citep{BerkPRL1996}. The prime denotes the derivative with respect
to $v$ while the superscript denotes the order in the wave amplitude
(proportional to $\left|\omega_{b}^{2}\right|/\nu^{2}$). Note that
$f_{0}{}^{(0)}$ corresponds to the equilibrium distribution $F_{0}$.
This ordering of the perturbation theory is satisfied for the entire
time evolution of the system so long as (i) steady asymptotic solutions
exist (guaranteed when $\alpha/\nu<0.96$ and $\nu>\nu_{\text{{crit}}}\approx2.05\left(\gamma_{L,0}-\gamma_{d}\right)$
\citep{Lilley2009PRL}) and (ii) the system is sufficiently close
to marginal stability, \textit{i.e. }$\gamma_{L,0}-\gamma_{d}\ll\gamma_{L,0}$
. When conditions (i) and (ii) are not satisfied, the mode may grow
to an amplitude that violates the $\left|\omega_{b}^{2}\right|\ll\nu^{2}$
assumption
. From Eq. \eqref{eq:KineticEq}, the $f_{l}$ satisfy
\begin{multline}
\frac{\partial f_{l}}{\partial t}+il\left(kv-\omega\right)f_{l}+\frac{\left(\omega_{b}^{2}f'_{l-1}+\omega_{b}^{2*}f'_{l+1}\right)}{2k}=C[f_{l}].\label{eq:KineticEqForN}
\end{multline}

In general, the solution of Eq. \eqref{eq:KineticEqForN} is an integral
over the time history of the system that involves delays in the argument
of $\omega_{b}\left(t\right)$. However, when stochastic collisions
dominate the system's dynamics, the system's memory is poorly retained
and the dynamics instead become essentially local in time \citep{duarte2018analytical,LestzDuartePoP2021}.
This occurs when $\nu\gg\gamma_{L,0}-\gamma_{d}$.
The constraints on these parameters become more restrictive as the
ratio of drag to scattering is increased \footnote{See Supplemental Material {[}url{]}, Secs. 6 - 8, which includes \citep{Lilley2009PRL,LestzDuartePoP2021}}.
Then, to first order in $\left|\omega_{b}^{2}\right|/\nu^{2}$ one
can disregard the time derivative in \eqref{eq:KineticEqForN}, yielding
\begin{equation}
f_{1}^{(1)}=-\frac{F'_{0}\omega_{b}^{2}\left(t\right)}{2k\nu}\int_{0}^{\infty}dse^{-i\left(\frac{kv-\omega}{\nu}\right)s-i\frac{\alpha^{2}}{\nu^{2}}s^{2}/2-s^{3}/3}.\label{eq:f1scatt}
\end{equation}

Using the reality rule $f_{-1}^{(1)}=f_{1}^{(1)*}$, Eq. \eqref{eq:KineticEqForN}
can be written to second order in $\left|\omega_{b}^{2}\right|/\nu^{2}$,
\begin{equation}
\frac{\partial f_{0}^{(2)}}{\partial t}+\frac{1}{2k}\left(\omega_{b}^{2}\left[f_{1}'^{(1)}\right]^{*}+\omega_{b}^{2*}f_{1}'^{(1)}\right)=C\left[f_{0}^{(2)}\right].\label{eq:f0K}
\end{equation}

Substitution of Eq. \eqref{eq:f1scatt} into Eq. \eqref{eq:f0K} shows
that only the spatially averaged distribution $f\left(v,t\right)\equiv\left(k/2\pi\right)\int_{0}^{2\pi/k}f\left(x,v,t\right)dx=F_{0}\left(v\right)+f_{0}^{(2)}\left(v,t\right)$
explicitly appears in the wave-particle power exchange to second order
in $\left|\omega_{b}^{2}\right|/\nu^{2}$. Since $\partial F_{0}/\partial t=0$
and $\left|F_{0}'\right|\gg\left|f_{0}'^{(2)}\right|$, one then obtains
from Eq. \eqref{eq:f0K} that the resonant particle response is regulated
by a QL diffusion-advection transport equation
\begin{multline}
\frac{\partial f\left(v,t\right)}{\partial t}-\frac{\partial}{\partial v}\left[\frac{\pi}{2k^{3}}\left|\omega_{b}^{2}\right|^{2}\mathcal{R}\left(v\right)\frac{\partial f}{\partial v}\right]=C[f-F_{0}],\label{eq:QLEq}
\end{multline}
with the resonance function given by

\begin{equation}
\mathcal{R}\left(v\right)=\frac{k}{\pi\nu}\int_{0}^{\infty}ds\:\cos\left(\frac{\left(kv-\omega\right)}{\nu}s+\frac{\alpha^{2}}{\nu^{2}}\frac{s^{2}}{2}\right)e^{-s^{3}/3}.\label{eq:WscattNdrag}
\end{equation}
$\mathcal{R}\left(v\right)$ gives the velocity-dependent strength
of the resonant wave-particle interaction. In the absence of collisions,
$\mathcal{R}\left(v\right)=\delta\left(v-\omega/k\right)$ would describe
an exact, unbroadened resonance. For all values of $\alpha/\nu$,
the property $\int_{-\infty}^{\infty}\mathcal{R}(v)dv=1$ is exactly
satisfied. The complete set of equations describing the QL system
is then given by Eqs. \eqref{eq:QLEq} and \eqref{eq:WscattNdrag},
coupled with the amplitude evolution equation 
\begin{equation}
\frac{d\left|\omega_{b}^{2}\left(t\right)\right|^{2}}{dt}=2\left(\gamma_{L}\left(t\right)-\gamma_{d}\right)\left|\omega_{b}^{2}\left(t\right)\right|^{2},\label{eq:domegabdt}
\end{equation}

where 

\begin{equation}
\gamma\left(t\right)=\frac{2\pi^{2}q^{2}\omega}{mk^{2}}\int_{-\infty}^{\infty}dv\mathcal{R}\left(v\right)\frac{\partial f\left(v,t\right)}{\partial v}.\label{eq:GrowthRate}
\end{equation}
A first-principles derivation of Eqs. \eqref{eq:domegabdt} and \eqref{eq:GrowthRate}
is shown in the Supplemental Material \footnote{See Supplemental Material {[}url{]}, Secs. 1 and 3, which includes
\citep{BerkPRL1996,duarte2018analytical,LestzDuartePoP2021}}.

Remarkably, a QL transport theory has emerged spontaneously from the
nonlinear one under the assumption that stochasticity dominates over
the relaxation timescale $(\nu\gg\gamma_{L,0}-\gamma_{d})$, even
in the presence of coherent drag which acts to preserve phase space
correlations. Furthermore, the QL system was obtained for an isolated,
discrete resonance without invoking any overlapping condition.

\begin{figure}[t]
\begin{centering}
\par\end{centering}
\includegraphics[width=1\columnwidth]{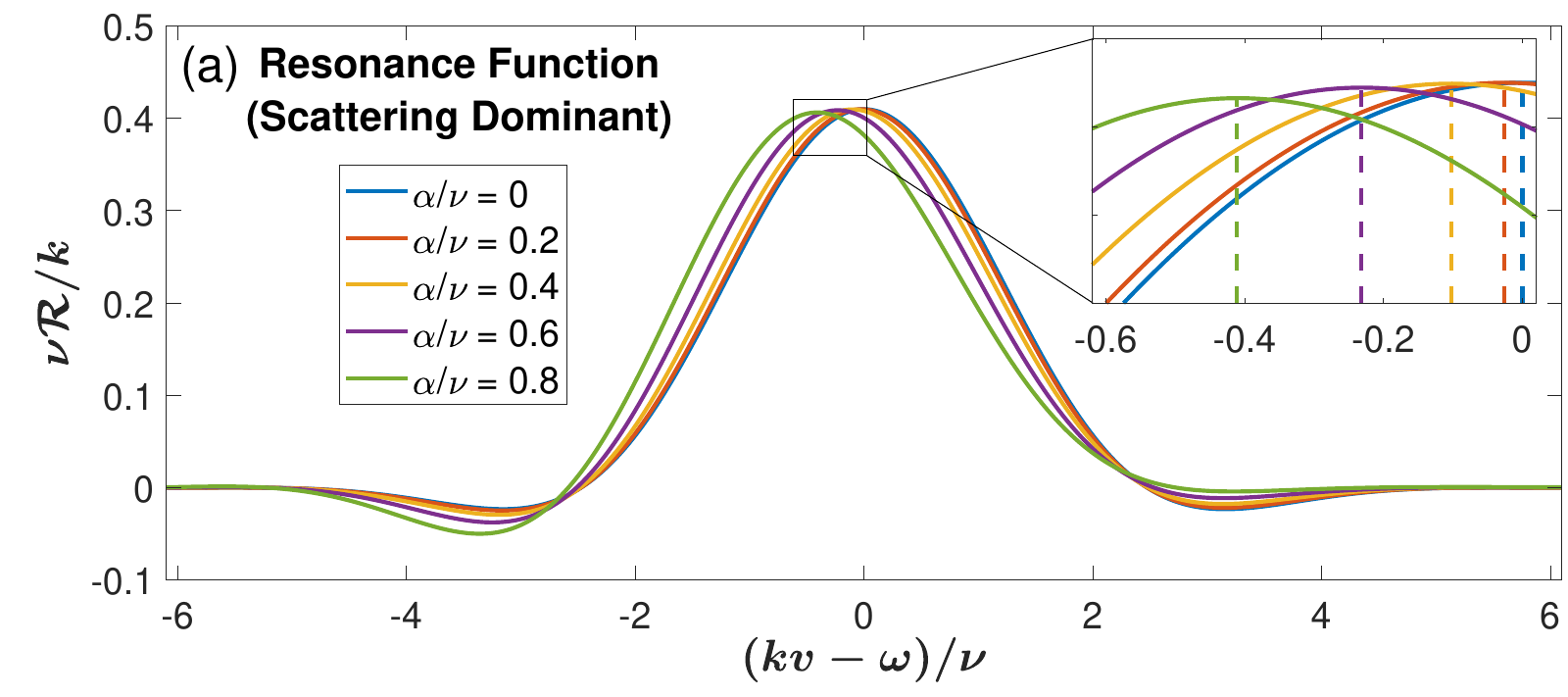}

\includegraphics[width=1\columnwidth]{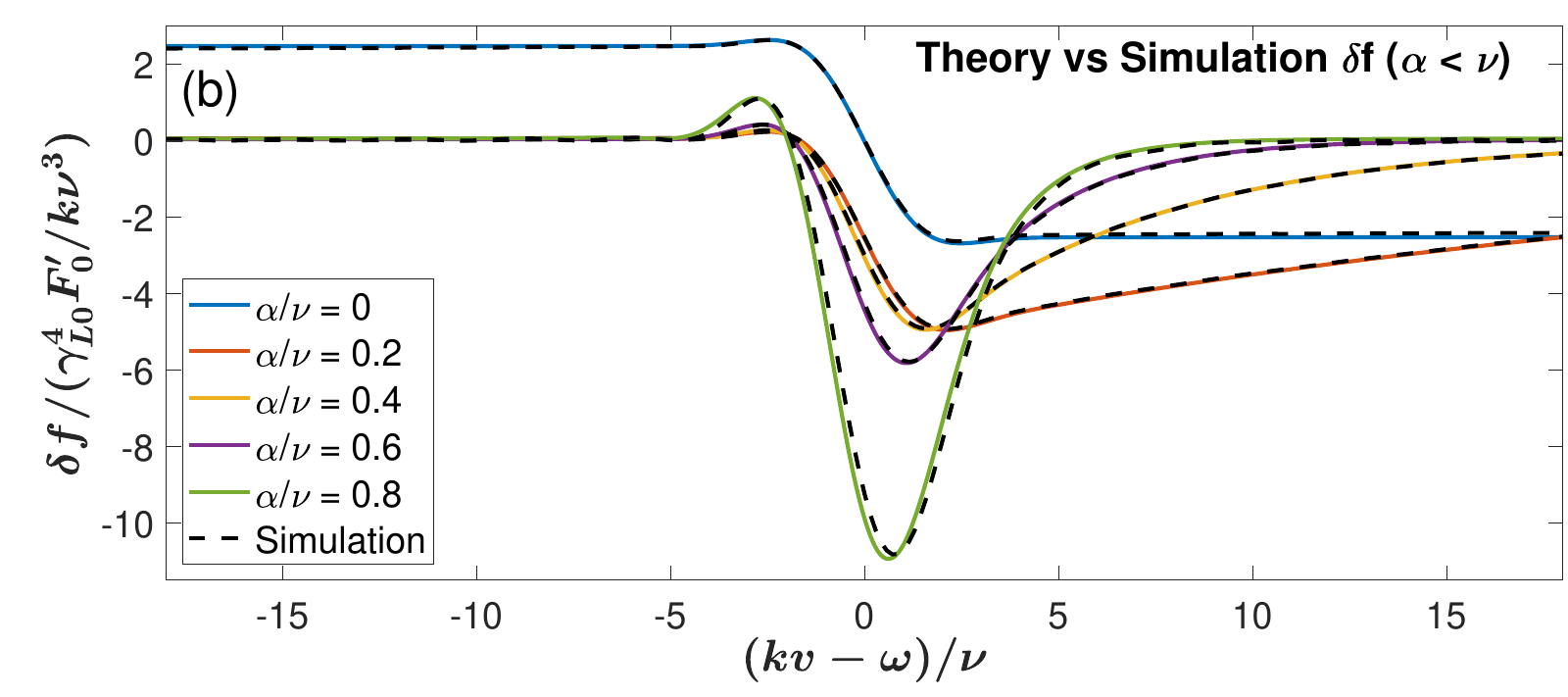}

\includegraphics[width=1\columnwidth]{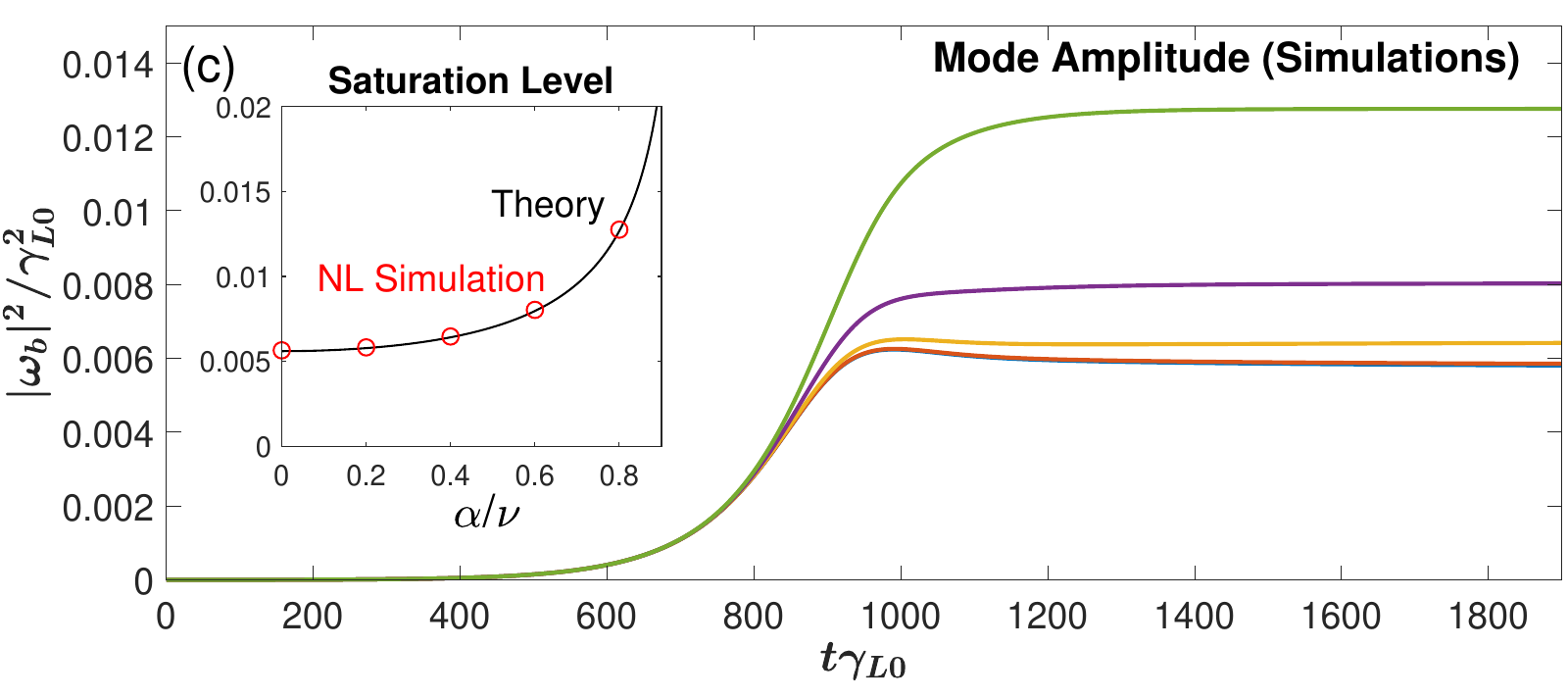}
\centering{}\caption{(a) Resonance function (Eq. (\ref{eq:WscattNdrag})), (b) saturated
distribution modification with respect to the initial equilibrium
distribution, $\delta f=f-F_{0}$ (Eq. \eqref{eq:deltafIntegrated}),
and (c) time evolution of the mode amplitude (proportional to $\left|\omega_{b}^{2}\right|$).
In (a) and (b), solid curves represent analytic expressions while
dashed curves are simulation results from the nonlinear Vlasov code
BOT using $\nu/(\gamma_{L,0}-\gamma_{d})$ = 20 and $\gamma_{d}/\gamma_{L,0}=0.99$.
Colored curves in (c) are simulation results. The inset plots the
saturation level from these simulations against the analytic prediction
\cite{Note5}. \label{FigWindFunc}}
\end{figure}
\textit{Shift of the resonance function due to symmetry breaking}.\textemdash{}
Plotting the resonance function $\mathcal{R}(v)$ for different values
of $\alpha/\nu$, as done in Fig. \ref{FigWindFunc}(a), demonstrates
that drag shifts the location of the strongest wave-particle interaction
in phase space. 
Moreover, drag acts to increase the strength of the interaction downstream
of the resonance and diminish it upstream. Although the quantitative
changes in the resonance function due to drag are seemingly small,
this fundamental change in the character of the wave-particle resonance
has substantial consequences for how the energetic particles are redistributed.
This shift is not analogous to a simple Doppler shift, but rather
the result of asymmetry present in the collisional dynamics due to
drag, which has a preferred direction. Without drag, $\mathcal{\mathcal{R}}(v)$
is perfectly symmetric about $kv-\omega=0$.

The shift of the resonance function implies a previously unrecognized
collisional modification to the resonance condition, which can be
obtained by calculating the location of the peak of $\mathcal{\mathcal{R}}(v)$,
leading to $\int_{0}^{\infty}\sin\left(s\left(\frac{kv-\omega}{\nu}\right)_{\text{{peak}}}+\frac{\alpha^{2}}{\nu^{2}}\frac{s^{2}}{2}\right)s\exp(-s^{3}/3)ds=0$.
Noting that only small $s$ contributes due to the strongly decaying
cubic exponential, the drag-modified resonance condition becomes $\left(kv-\omega\right)_{\text{{peak}}}\approx-3^{1/3}\Gamma\left(4/3\right)\alpha^{2}/2\nu$,
which is accurate to within $1.5\%$ for $\alpha/\nu<1$. Shifted resonance lines, captured here using drag in a considerably
reduced theory, are also known to exist in the strong turbulence framework
\citep{BirminghamShift1972} as a result of turbulence modification
of ensemble average orbits.

\textit{Particle relaxation}.\textemdash The derived QL system also
reproduces key features of the complete nonlinear system, namely the
perturbed distribution function ($\delta f\equiv f-F_{0})$ and wave
saturation amplitude. When stochasticity regulates the timescale for
the mode growth, 
Eq. \eqref{eq:QLEq} can be further simplified to $\left(\nu^{3}/k\right)\delta f'+\alpha^{2}\delta f=-\left(\pi\left|\omega_{b}^{2}\right|^{2}/2k^{2}\right)\mathcal{R}\left(v\right)F'_{0}$,
which can be solved for by direct integration:
\begin{widetext}
\begin{multline}
\delta f\left(v,t\right)=\frac{\left|\omega_{b}^{2}\left(t\right)\right|^{2}F'_{0}}{2k\nu^{3}}\left\{ c\left(\frac{\alpha}{\nu}\right)-\int_{0}^{\infty}\frac{ds\;e^{-s^{3}/3}}{\alpha^{4}/\nu^{4}+s^{2}}\left[\frac{\alpha^{2}}{\nu^{2}}\cos\left(\frac{\left(kv-\omega\right)s}{\nu}+\frac{\alpha^{2}}{\nu^{2}}\frac{s^{2}}{2}\right)+s\sin\left(\frac{\left(kv-\omega\right)s}{\nu}+\frac{\alpha^{2}}{\nu^{2}}\frac{s^{2}}{2}\right)\right]\right\} .\label{eq:deltafIntegrated}
\end{multline}
\end{widetext}

The integration constant $c(\alpha/\nu)$ is determined by enforcing
particle conservation: $\int_{-v_{\text{max}}}^{v_{\text{max}}}\delta fdv=0$.
The velocity structure of $\delta f$ does not evolve in time, as
all time dependence is contained in the overall factor $\left|\omega_{b}^{2}\left(t\right)\right|^{2}$.
The relaxed distribution is plotted in Fig. \ref{FigWindFunc}(b)
for several values of $\alpha/\nu<1$. Remarkably, a small quantitative
asymmetry in the collisional dynamics due to drag can have a large
qualitative effect on the saturated distribution, even though the
corresponding changes in the resonance function are less dramatic.
When no drag is present $(\alpha=0)$, $\delta f$ is antisymmetric
with constant plateaus outside of a narrow transition region near
the peak of the resonance. As the ratio of drag to scattering is increased,
the plateau upstream of the resonance $(kv>\omega)$ instead decays
in velocity space at a rate proportional to $\alpha^{2}/\nu^{2}$. Particle conservation shifts the entire distribution downards once
the symmetry is broken, eliminating nearly all of the downstream particle
redistribution, even for very small amounts of drag. The drag-induced
modifications to $\delta f$ were compared against simulations performed
with the nonlinear 1D Vlasov code BOT, which solves the plasma kinetic
equation (Eq. \eqref{eq:KineticEq}) directly \citep{Lilley2010,BOTcode}.
The dashed black curves in Fig. \ref{FigWindFunc}(b) show the simulation
results for each value of $\alpha/\nu$, demonstrating excellent agreement.

\textit{Instability saturation level}.\textemdash Drag has a destabilizing
effect on the underlying instability, leading the wave to saturate
at a larger amplitude with increasing $\alpha/\nu$. An analytically tractable example is the first order correction to
the saturation amplitude due to drag. Substituting Eq. \eqref{eq:deltafIntegrated}
into Eq. \eqref{eq:GrowthRate}, to lowest order in $\alpha^{2}/\nu^{2}$,
one finds (the details of the derivation are given in the Supplemental
Material \footnote{See Supplemental Material {[}url{]}, Sec. 2, which includes \citep{DuarteAxivPRL,duarte2019collisional,LestzDuartePoP2021}})
\begin{equation}
\gamma_{L}\left(t\right)\simeq\gamma_{L,0}\left\{ 1-\frac{\left|\omega_{b}^{2}\left(t\right)\right|^{2}}{2\nu^{4}}\left[\Gamma\left(\frac{1}{3}\right)\left(\frac{3}{2}\right)^{1/3}\frac{1}{3}-\frac{\pi}{2}\frac{\alpha^{2}}{\nu^{2}}\right]\right\} .\label{eq:GammaL}
\end{equation}
At saturation, \textit{i.e.}, when $\gamma_{L}\left(t\right)=\gamma_{d}$,
then the mode amplitude $\hat{E}$ is given by
\begin{equation}
\left|\omega_{b,\text{{sat}}}\right|=\sqrt{\frac{qk\hat{E}_{\text{sat}}}{m}}\simeq\left[\frac{2\left(1-\gamma_{d}/\gamma_{L,0}\right)}{\Gamma\left(\frac{1}{3}\right)\left(\frac{3}{2}\right)^{1/3}\frac{1}{3}-\frac{\pi}{2}\frac{\alpha^{2}}{\nu^{2}}}\right]^{1/4}\nu,\label{eq:SatLevel}
\end{equation}
which is the same as calculated directly from nonlinear theory (Ref.
\citep{LestzDuartePoP2021}). In the absence of drag, the saturation
in Eq. \eqref{eq:SatLevel} recovers the level calculated in Ref.
\citep{BerkPPR1997}. While not possible to express in terms of elementary
functions, it can nonetheless be proven analytically that the QL saturation
level also reproduces the fully nonlinear one whenever a steady state
solution exists. 

Fig \ref{FigWindFunc}(c) shows the time evolution of the mode amplitude
in fully nonlinear simulations. The saturation levels are in excellent
agreement with the unapproximated analytic expression \footnote{See Supplemental Material {[}url{]}, Sec. 6, which includes \citep{Lilley2009PRL,LestzDuartePoP2021}},
and demonstrate that even moderate amounts of drag can appreciably
increase the saturated mode amplitude. For example, $\alpha/\nu=0.6$
yields a 40\% increase in the saturated amplitude relative to if drag
were neglected. Consequently, this increase in mode amplitude enhances
the resonant particle transport, as shown in Fig. \ref{FigWindFunc}(b).
As $\alpha/\nu$ is increased from 0.6 to 0.8, the magnitude of $\delta f$
nearly doubles. 

\textit{Resonance splitting due to large drag}.\textemdash It is also
interesting to consider the structure of the resonance when drag dominates
over scattering, corresponding to instabilities which will eventually
reach a strongly nonlinear regime, with the potential for substantial
transport. Although for $\alpha/\nu>0.96$ no steady state solution
is allowed within the near-threshold perturbation theory \citep{Lilley2009PRL},
all of the derivations to this point nonetheless remain valid during
the early growth phase, up until the mode amplitude exceeds the assumed
$\left|\omega_{b}^{2}\right|/\nu^{2}\ll1$ ordering. 

\begin{figure}[t]
\begin{centering}
\par\end{centering}
\includegraphics[width=1\columnwidth]{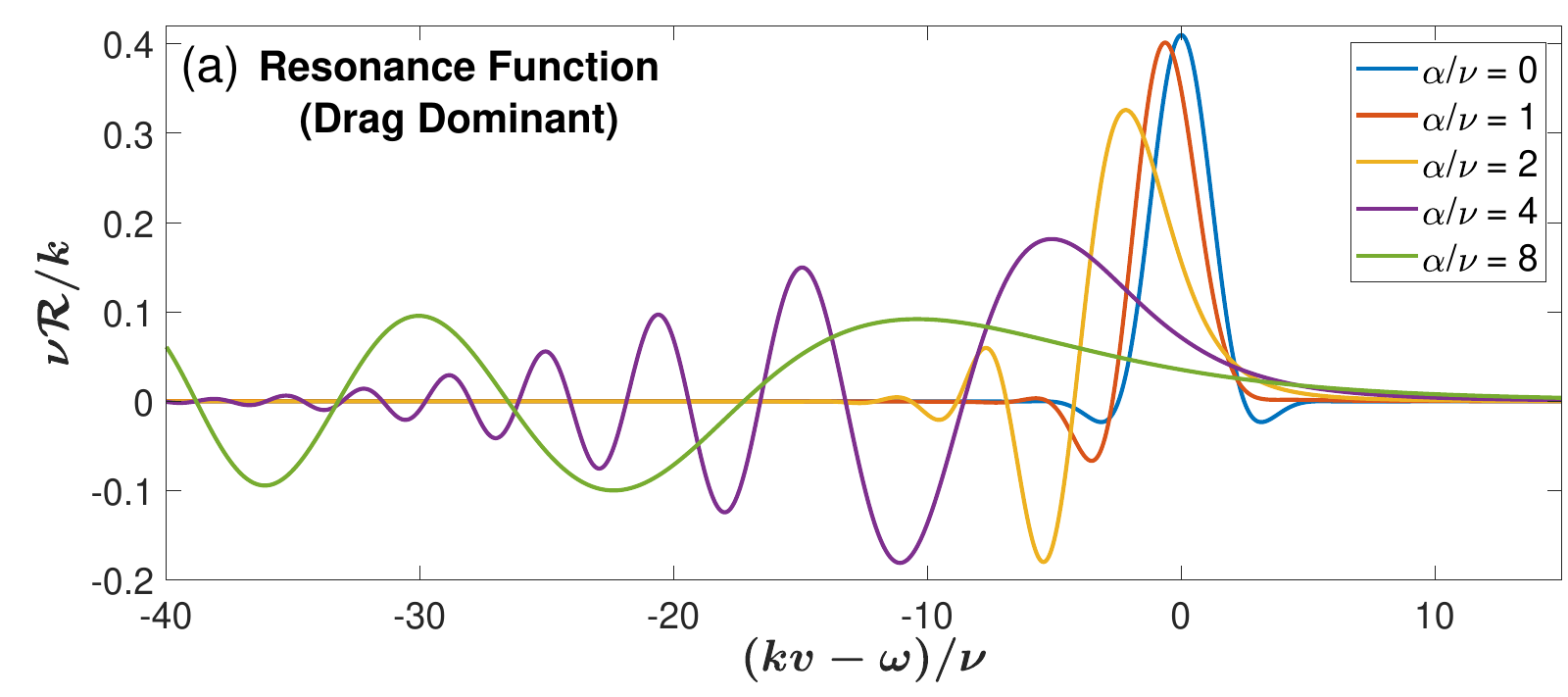}

\includegraphics[width=1\columnwidth]{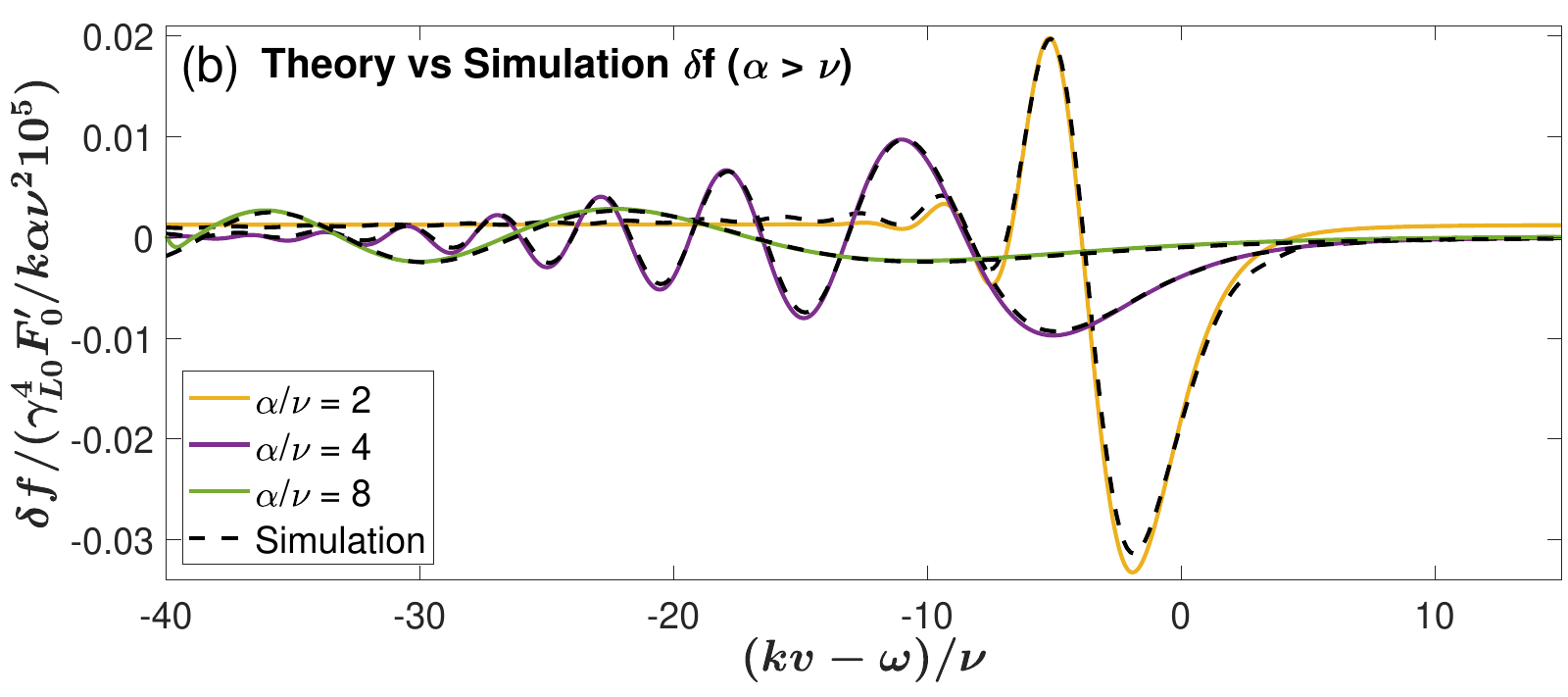}
\centering{}\caption{(a) Resonance function (Eq. (\ref{eq:WscattNdrag})) and (b) saturated
distribution modification with respect to the initial equilibrium
distribution, $\delta f=f-F_{0}$. The dashed curves in part (b) were
produced at $t\gamma_{L,0}=300$ from nonlinear BOT simulations with
$\nu/(\gamma_{L,0}-\gamma_{d})=20$, $\gamma_{d}/\gamma_{L,0}=0.99,$
and initial condition $\omega_{b}/\gamma_{L,0}=10^{-8}$, while the
solid curves represent Eq. \eqref{eq:deltafIntegrated}.\label{FigWindFunc-2} }
\end{figure}

For $\alpha\gg\nu$, a pronounced splitting of the resonance function
occurs, as shown in Fig. \ref{FigWindFunc-2}(a). Simultaneously, the resonance
function broadens beyond its original width, proportional to $\nu$,
instead becoming proportional to $\alpha$. It can be shown from Eq.
\eqref{eq:WscattNdrag} that for $kv<\omega$ in the limit of $(kv-\omega)^{2}\gg\alpha^{2}\gg\nu^{2}\gg\gamma^{2}$,
the resonance function has the following asymptotic behavior:

\begin{equation}
\mathcal{R}(v)=\frac{1}{\alpha}\sqrt{\frac{2}{\pi}}\exp\left[\frac{1}{3}\left(\frac{\nu}{\alpha}\frac{kv-\omega}{\alpha}\right)^{3}\right]\cos\left[\frac{(kv-\omega)^{2}}{2\alpha^{2}}-\frac{\pi}{4}\right].\label{eq:rwinbig}
\end{equation}
 Hence the resonance function's extrema are given by $\left(kv-\omega\right)_{\text{crit}}=-\alpha\sqrt{2\pi}\sqrt{n+1/4}$
for $n\geq0$. In particular, $n=0$ gives the central shift in the
resonance function, showing that it becomes proportional to $\alpha$
when drag dominates instead of $\alpha^{2}$ when scattering dominates.
Strictly speaking, similar extrema are present even when $\alpha\ll\nu$, however they are substantially smaller than the primary peak.

The splitting of the resonance function is a novel quasilinear effect
not previously identified in plasmas. This behavior implies that for
sufficiently large drag there can be multiple regions of phase space
where particles are efficiently interacting with the wave, even for
the idealized case of a monochromatic wave with a uniform background.

During the early growth phase in the drag-dominated regime, $\delta f$
has identical asymptotic behavior as the resonance function: $\delta f(v)\propto-\mathcal{R}(v)$,
inheriting the same pattern of decaying oscillations. Strong agreement is found between the analytic $\delta f$ calculated
with Eq. \ref{eq:deltafIntegrated} and a nonlinear BOT simulation
with $\alpha\gg\nu$, as shown in Fig. \ref{FigWindFunc-2}(b) for
the cases with clear splitting. Consequently, large amounts of drag
act to extend the downstream region of phase space where significant
transport occurs. 

\textit{Relevance to fusion plasmas}.\textemdash The effects of drag
derived in this Letter are most prominent when drag approaches or
exceeds scattering within a narrow resonance
. A heuristic scaling useful for guiding intuition is given by $\alpha/\nu\sim\mathcal{E_{\text{{res}}}}^{1/2}n_{e}^{1/6}/T_{e}^{3/4}\omega^{1/6}$
\citep{LestzDuartePoP2021}. Quantitatively, $\alpha/\nu$ can be
numerically calculated with rigor in fusion plasmas using a kinetic
equilibrium reconstruction, solving for the eigenmode structure with
an MHD code, following realistic orbits, and averaging over resonant
surfaces. This method is outlined in the Supplemental Material \footnote{See Supplemental Material {[}url{]}, Secs. 4 and 5, which includes
\citep{BerkPPR1997,DuarteAxivPRL,DuarteNF2018Likelihood,ZhangLinChen2008PRL,LangFu2011,KayeNF2007,ZhangHeidbrinkLAPD_PoP2099,Heidbrink_LAPD_2012,Gekelman_LAPD_2016,SarfatySkiffPRL1998,SchroderSkiff2017}} and was previously applied to NSTX, DIII-D, and ITER \citep{DuarteAxivPRL,DuartePoP2017,DuarteNF2018Likelihood},
including the effect of enhanced diffusion due to microturbulence
\citep{LangFu2011,ZhangLinChen2008PRL}. 

An increase in $\alpha/\nu$ due to the reduction of turbulence has
been previously identified as a reliable indicator of the onset of
chirping behavior for \Alfvenic modes \citep{DuarteAxivPRL}, which
is ubiquitous in spherical tokamaks but rarely observed in conventional
tokamaks. Turbulence is typically lower in spherical tokamaks such
as NSTX due to favorable curvature and enhanced rotation shear \citep{KayeNF2007},
allowing $\alpha\sim\nu$. Similar findings exist on other devices.
Calculations by Lilley \textit{et al}. found that $\alpha/\nu=0.6-5$
at the TAE resonance for beam-heated MAST plasmas \citep{Lilley2009PRL},
with the practical consequence of explaining why those discharges
often feature bursting TAEs, in contrast to radio-frequency-heated
discharges with larger scattering rates. In addition, Lesur \textit{et
al}. analyzed an uncommon case of chirping in JT-60U by fitting the
observed frequency sweeping to a theoretical model, enabling the extraction
of the collisional coefficients, finding $\alpha/\nu$ in the range
of $0.2-0.7$ \textendash{} consistent with measured plasma parameters
\citep{Lesur2010EspectrDetermination}. Furthermore, simulations of
an ITPA tokamak benchmark case and the W7-X stellarator using a global
hybrid MHD-kinetic code performed by Slaby \textit{et al.} concluded
that including realistic amounts of drag in the Fokker-Planck collision
operator affected both the AE saturation level and its long term nonlinear
behavior \citep{Slaby2019NF}. Lastly, negative triangularity experiments
on DIII-D observed a greater tendency for chirping than in matched
positive triangularity discharges \citep{Van_Zeeland_2019}. Van Zeeland
\textit{et al}. attributed this difference to the lower level of turbulent
scattering in negative triangularity, increasing $\alpha/\nu$ and
making non-steady behavior more likely according to theory. The relative
propensity for chirping in several distinct scenarios empirically
supports the relevance of drag to wave-particle dynamics in fusion
plasmas, especially tokamaks with low aspect ratio or negative triangularity. 

\textit{Connection to galactic dynamics}.\textemdash Beyond fusion
plasmas, the methods presented in this Letter are directly applicable
to the resonant gravitational interaction of the rotating galactic
bar and heavy bodies in its orbit such as black holes, massive astrophysical
compact halo objects, and stars in a tepid disk \citep{Fouvry2015}.
As recently demonstrated by Hamilton et al. \citep{Hamilton2022},
the collisional bump-on-tail problem in plasmas (studied in this work)
can be made isomorphic to this application. In this formalism, the
role of the AE is played by the rigidly rotating galactic bar, the
resonant fast ions are replaced by the orbiting bodies, and the electrostatic
potential is replaced by a gravitational one. Diffusive scattering
occurs due to stochastic potential fluctuations, while heavy bodies
experience drag when passing through a background of lighter masses.
The resonant interaction exerts a torque on the bar and leaves a collisionally-dependent
imprint on the spatial distribution of surrounding bodies in the galaxy,
which in tandem with observations can be used to constrain theoretical
models of dark matter. Specifically, the collisions experienced by
orbiting heavy compact objects are dominated by drag when their mass
ratio to the background particles is sufficiently large \citep{BinneyTremaine,Ciotti2021}. Hence, the novel influence of
drag on wave-particle interactions derived in this Letter is relevant
for an accurate description of the resonant galactic bar-heavy body
system in the presence of collisions. 

\begin{acknowledgments}
This work was supported by the US Department of Energy under contracts
DE-AC02-09CH11466, DE-SC0020337, and DE-FC02-04ER54698. VND and JBL
thank H. L. Berk for multiple clarifying discussions pertaining to
nonlinear kinetic instabilities, M. K. Lilley for making the code
BOT openly available as well as for clarifying comments regarding
the boundary conditions employed there, C. Hamilton for enlightening
discussions on the link between plasma physics and galactic dynamics,
and to L. Comisso, V. Skoutnev, W. W. Heidbrink, and R. Nazikian for
helpful feedback.

\end{acknowledgments}

\bibliographystyle{apsrev4-2}
\addcontentsline{toc}{section}{\refname}

\end{document}

\title{Supplemental Material of the Letter \\ ``\textit{Shifting and splitting
of resonance lines due to dynamical friction in plasmas}''}
\author{V. N. Duarte}
\thanks{\authnote}
%\email{vduarte@pppl.gov}
\address{Princeton Plasma Physics Laboratory, Princeton University, Princeton,
NJ, 08543, USA}

\author{J. B. Lestz}
\thanks{\authnote}
%\email{lestzj@fusion.gat.com}
\address{General Atomics, San Diego, CA, 92121, USA}
\address{Department of Physics and Astronomy, University of California, Irvine,
CA, 92697, USA}

\author{N. N. Gorelenkov}
\address{Princeton Plasma Physics Laboratory, Princeton University, Princeton,
NJ, 08543, USA}
\author{R. B. White}
\address{Princeton Plasma Physics Laboratory, Princeton University, Princeton,
NJ, 08543, USA}
\maketitle

\section{Derivation of the evolution equation, the growth rate and the relaxed
distribution}

The power transfer from the resonant particles satisfying $kv\approx\omega$
to the wave is
\[
P=-\frac{q\omega}{k}\int dv\int dxE^{*}\left(x,t\right)f\left(x,v,t\right).
\]
The power balance equation, for a constant background damping rate
$\gamma_{d}$, is

\begin{equation}
\frac{d}{dt}\int dx\frac{\left|E\left(x,t\right)\right|^{2}}{8\pi}=P-\gamma_{d}\int dx\frac{\left|E\left(x,t\right)\right|^{2}}{8\pi}.\label{eq:PowBal}
\end{equation}
As in the Letter, the distribution is considered in the form $f\left(x,v,t\right)=f_{0}\left(v,t\right)+\underset{l=1}{\overset{\infty}{\sum}}\left(f_{l}\left(v,t\right)e^{il\left(kx-\omega t\right)}+c.c.\right)$,
with the small parameter $\left|\omega_{b}^{2}\right|/\nu^{2}\ll1$
leading to the ordering $\left|f_{0}'^{(0)}\right|\gg\left|f_{1}'^{(1)}\right|\gg\left|f_{0}'^{(2)}\right|,\left|f_{2}'^{(2)}\right|$
\citep{BerkPRL1996}. The prime denotes the derivative with respect
to $v$ while the superscript denotes the order in the wave amplitude
(proportional to $\left|\omega_{b}^{2}\right|/\nu^{2}$). Performing
the spatial integration of Eq. \eqref{eq:PowBal} over a wavelength
leads to
\begin{equation}
\frac{d\hat{E}\left(t\right)}{dt}=-\frac{4\pi q\omega}{k}\int dvf_{1}\left(v,t\right)-\gamma_{d}\hat{E}\left(t\right)\label{eq:MasterPowerBalance}
\end{equation}
where $q$ is the energetic particle electric charge. Consider the
above equation up to cubic order in the electric field amplitude,
i. e., $f_{1}\simeq f_{1}^{(1)}+f_{1}^{(3)}$ ($f_{1}^{(2)}=0$ because
the structure of the kinetic equation {[}Eq. (2) of the Letter{]}
does not allow for a first-order Fourier $f_{1}$ to have even powers
in the electric field amplitude), where $f_{1}^{(1)}$ and $f_{1}^{(3)}$
can be found by direct integration of the kinetic equation (the details
are given in Sec. III of this Supplemental Material)

\begin{equation}
f_{1}^{(1)}=-\frac{\omega_{b}^{2}\left(t\right)}{2k\nu}\frac{\partial F_{0}}{\partial v}\int_{0}^{\infty}dse^{-i\left(\frac{kv-\omega}{\nu}\right)s-i\frac{\alpha^{2}}{\nu^{2}}s^{2}/2-s^{3}/3}\label{eq:f11}
\end{equation}
and
\begin{equation}
f_{1}^{(3)}=-\frac{\omega_{b}^{2}\left(t\right)}{2k\nu}\frac{\partial f_{0}^{(2)}}{\partial v}\int_{0}^{\infty}dse^{-i\left(\frac{kv-\omega}{\nu}\right)s-i\frac{\alpha^{2}}{\nu^{2}}s^{2}/2-s^{3}/3}.\label{eq:f13}
\end{equation}
Using the definition $\omega_{b}^{2}\equiv qk\hat{E}/m$ and Eqs.
\eqref{eq:f11} and \eqref{eq:f13}, Eq. \eqref{eq:MasterPowerBalance}
becomes

\begin{multline}
\frac{d\omega_{b}^{2}\left(t\right)}{dt}=-\gamma_{d}\omega_{b}^{2}\left(t\right)+\frac{2\pi q^{2}\omega}{k\nu m}\omega_{b}^{2}\left(t\right)\times\\
\times\int_{-\infty}^{\infty}dv\left(\frac{\partial F_{0}}{\partial v}+\frac{\partial f_{0}^{(2)}}{\partial v}\right)\int_{0}^{\infty}dse^{-i\left(\frac{kv-\omega}{\nu}\right)s-i\frac{\alpha^{2}}{\nu^{2}}s^{2}/2-s^{3}/3}.\label{eq:MasterPowerBalance-1-1}
\end{multline}
The equilibrium distribution can be approximated by a constant slope
within narrow resonance layers. Since $\partial F_{0}/\partial v\gg\partial f_{0}^{(2)}/\partial v$
and 
\[
\int dv\int_{0}^{\infty}ds\sin\left[\left(\frac{kv-\omega}{\nu}\right)s+\frac{\alpha^{2}}{\nu^{2}}\frac{s^{2}}{2}\right]e^{-s^{3}/3}=0,
\]
then the power balance equation {[}Eq. \eqref{eq:MasterPowerBalance}{]}
leads to the amplitude evolution equation 
\begin{equation}
\frac{d\left|\omega_{b}^{2}\left(t\right)\right|^{2}}{dt}=2\left(\gamma_{L}\left(t\right)-\gamma_{d}\right)\left|\omega_{b}^{2}\left(t\right)\right|^{2}\label{eq:domegabdt}
\end{equation}
with the growth rate $\gamma_{L}\left(t\right)$ given by 
\begin{equation}
\gamma_{L}\left(t\right)=\frac{2\pi^{2}q^{2}\omega}{mk^{2}}\int_{-\infty}^{\infty}dv\mathcal{R}\left(v\right)\frac{\partial f\left(v,t\right)}{\partial v},\label{eq:GrowthRate}
\end{equation}
where the spatially-integrated distribution used in quasilinear theory
is $f\left(v,t\right)\simeq F_{0}\left(v\right)+f_{0}^{(2)}\left(v,t\right)$.
At $t=0$, $\gamma_{L}=\gamma_{L,0}=\left(2\pi^{2}q^{2}\omega/mk^{2}\right)\partial F_{0}/\partial v$.
Eqs. \eqref{eq:domegabdt} and \eqref{eq:GrowthRate}, together with
the quasilinear equation for the distribution relaxation (Eq. (5)
of the Letter, derived for a marginally unstable mode)

\begin{multline}
\frac{\partial f\left(v,t\right)}{\partial t}-\frac{\partial}{\partial v}\left[\frac{\pi}{2k^{3}}\left|\omega_{b}^{2}\right|^{2}\mathcal{R}\left(v\right)\frac{\partial f\left(v,t\right)}{\partial v}\right]=\\
=\frac{\nu^{3}}{k^{2}}\frac{\partial^{2}\left(f\left(v,t\right)-F_{0}\right)}{\partial v^{2}}+\frac{\alpha^{2}}{k}\frac{\partial\left(f\left(v,t\right)-F_{0}\right)}{\partial v}\label{eq:QLEq}
\end{multline}
constitute the full set of transport equations. In Eq. \eqref{eq:QLEq},
the resonance function is given by
\begin{equation}
\mathcal{R}\left(v\right)=\frac{k}{\pi\nu}\int_{0}^{\infty}ds\:\cos\left(\frac{\left(kv-\omega\right)}{\nu}s+\frac{\alpha^{2}}{\nu^{2}}\frac{s^{2}}{2}\right)e^{-s^{3}/3}.\label{eq:WscattNdrag}
\end{equation}
$\mathcal{R}\left(v\right)$ gives the velocity-dependent strength
of the resonant wave-particle interaction. The full dynamic solution
of Eq. \eqref{eq:QLEq} for $\delta f\equiv f\left(v,t\right)-F_{0}\left(v\right)$
is 
\begin{multline*}
\delta f=\frac{F'_{0}}{2k^{2}\nu}\int_{0}^{t}dt_{1}\left|\omega_{b}^{2}\left(t_{1}\right)\right|^{2}\frac{\partial}{\partial v}\int_{0}^{\infty}dse^{-s^{3}/3}e^{-\nu s^{2}\left(t-t_{1}\right)}\times\\
\times\cos\left[\frac{kv-\omega}{\nu}s+\frac{\alpha^{2}s}{\nu}\left(\frac{s}{2\nu}+t-t_{1}\right)\right]+\text{{const.}}
\end{multline*}
Quasi-steady solutions can be obtained by neglecting the time derivative
in Eq. \eqref{eq:QLEq}. The solution for such system is
\begin{multline}
\delta f\left(v,t\right)=\frac{\pi}{2}\frac{\left|\omega_{b}^{2}\left(t\right)\right|^{2}}{k\nu^{3}}F'_{0}\Biggl\{ c\left(\frac{\alpha}{\nu}\right)-\\
\left.-\int^{v}dv'\mathcal{R}\left(v'\right)\exp\left[-\frac{\alpha^{2}}{\nu^{3}}k\left(v-v'\right)\right]\right\} .\label{eq:deltaf}
\end{multline}

Using Eq. \eqref{eq:WscattNdrag}, the velocity integral in Eq. \eqref{eq:deltaf}
can be performed, which leads to the expression shown in Eq. (10)
of the Letter,
\begin{widetext}
\begin{multline*}
\delta f\left(v,t\right)=\frac{\left|\omega_{b}^{2}\left(t\right)\right|^{2}}{2k\nu^{3}}F'_{0}\left\{ c\left(\frac{\alpha}{\nu}\right)-\int_{0}^{\infty}\frac{ds\;e^{-s^{3}/3}}{\alpha^{4}/\nu^{4}+s^{2}}\left[\frac{\alpha^{2}}{\nu^{2}}\cos\left(\frac{\left(kv-\omega\right)s}{\nu}+\frac{\alpha^{2}}{\nu^{2}}\frac{s^{2}}{2}\right)+s\sin\left(\frac{\left(kv-\omega\right)s}{\nu}+\frac{\alpha^{2}}{\nu^{2}}\frac{s^{2}}{2}\right)\right]\right\} .
\end{multline*}
\end{widetext}

The integration constant $c(t,\alpha/\nu)$ is determined by enforcing
particle conservation: $\int_{-v_{\text{max}}}^{v_{\text{max}}}\delta fdv=0$.
A subtlety of the nonlinear simulations shown in the Letter's Fig.
1(b) is that there is a slight discrepancy for the special case of
no drag. The simulation uses a periodic/free-flowing boundary condition
in order to conserve particles, imposing $\delta f(v_{\text{{max}}})=\delta f(-v_{\text{{max}}})$,
while $\delta f$ is antisymmetric when $\alpha=0$. To satisfy both
conditions, $\delta f$ becomes pinned to zero in the simulations
at the boundary $(kv_{\text{{max}}}/\nu=1250\pi)$, in contrast to
Eq. \eqref{eq:deltaf}, which conserves particles without the additional
numerical constraint. This artificial boundary effect has only a minor
effect near the resonance (\textit{e.g.} the subdomain shown in Fig.
1(b)), which is the physical region of interest. Aside from this technical
subtlety, the simulations confirm that the derived QL system evolves
to the same particle distribution as the exact nonlinear one.

It is important to note that the coefficients $\nu$ and $\alpha$
are the effective scattering and drag collision frequencies, which
are enhanced with respect to the $90\lyxmathsym{\textdegree}$ pitch
angle scattering rate and the inverse slowing down time. The effective
frequencies represent the enhancement of collisional effects in narrow
resonance layers, even in weakly collisional plasmas \citep{SuOberman1968,BerkBreizmanPRL1992,Callen2014PoPCoulombCollsions}.
They scale as $\nu\sim\nu_{\perp}\omega^{2}/\left(k\Delta v\right)^{2}$
and $\alpha\sim\tau_{s}^{-1}\omega/\left(k\Delta v\right)$ where
$k\Delta v\sim\nu$ is the characteristic broadening width, $\nu_{\perp}$
is the $90$-degree pitch angle scattering rate and $\tau_{s}$ is
the slowing down time. Detailed expressions for $\nu$ and $\alpha$
in a tokamak are described by Eqs. 6 and 7 of Ref. \citep{DuartePoP2017}.
Due to the narrowness of the resonances, the resonant particle dynamics
can be strongly controlled by collisions even if their non-resonant
dynamics hardly feel their effect.

\section{Derivation of the growth rate to lowest order in $\alpha^{2}/\nu^{2}$}

The condition $\alpha\ll\nu$, common in scenarios of fast ion resonances
with \Alfven eigenmodes in tokamaks \citep{DuarteAxivPRL} further
allows the resonance function {[}Eq. \eqref{eq:WscattNdrag}{]} to
be approximately written as 
\begin{equation}
\mathcal{R}\left(v\right)\simeq\left[1-\frac{\alpha^{2}}{2\nu^{2}}\frac{kv-\omega}{\nu}\right]\mathcal{R}_{\text{{scatt}}}\left(v\right),\label{eq:WindFunctApprox}
\end{equation}
where 
\begin{equation}
\mathcal{R}_{\text{scatt}}\left(v\right)=\frac{k}{\pi\nu}\int_{0}^{\infty}ds\:\cos\left(\frac{kv-\omega}{\nu}s\right)e^{-s^{3}/3}\label{eq:WindFunctScatt}
\end{equation}
is the symmetric resonance function in the absence of drag \citep{duarte2019collisional}.
Substituting Eq. \eqref{eq:deltaf}, into Eq. \eqref{eq:GrowthRate},
one obtains

\begin{equation}
\begin{array}{c}
\gamma_{L}\left(t\right)=\gamma_{L0}\left\{ 1-\frac{\pi}{2}\frac{\left|\omega_{b}^{2}\left(t\right)\right|^{2}}{\nu^{4}}\left[k\int_{-\infty}^{\infty}dv\mathcal{R}^{2}\left(v\right)-\right.\right.\\
\left.\left.-\frac{\alpha^{2}}{\nu^{2}}k^{2}\int_{-\infty}^{\infty}dv\mathcal{R}\left(v\right)\int^{v}dv'\mathcal{R}\left(v'\right)\exp\left[-\frac{\alpha^{2}}{\nu^{3}}k\left(v-v'\right)\right]\right]\right\} 
\end{array}\label{eq:GammaLInterm-1}
\end{equation}
To first order in $\alpha^{2}/\nu^{2}$, the resonance function in
Eq. \eqref{eq:GammaLInterm-1} can be approximated by Eq. \eqref{eq:WindFunctApprox}.
The first term of Eq. \eqref{eq:GammaLInterm-1} can be simplified
by noting that because $\mathcal{R}_{\text{scatt}}\left(v\right)$
(given by Eq. \ref{eq:WindFunctScatt}) is an even function with respect
to $kv=\omega$, then $\int_{-\infty}^{\infty}\left(kv-\omega\right)\mathcal{R}_{\text{scatt}}^{2}dv=0$.
Therefore the first term only contributes via $\int_{-\infty}^{\infty}\mathcal{R}_{\text{scatt}}^{2}dv=\frac{2k}{\pi\nu}\left[\frac{1}{6}\Gamma\left(\frac{1}{3}\right)\left(\frac{3}{2}\right)^{1/3}\right]$.

For the second term of Eq. \eqref{eq:GammaLInterm-1}, it should be
noted that, because $v'\leq v$ in the inner integral, then the exponential
is always within the interval $(0,1]$. The term involving the double
integral in Eq. \eqref{eq:GammaLInterm-1} is already multiplied by
$\alpha^{2}/\nu^{2}$, so we seek to calculate the integrals to zeroth
order in this small parameter.

Then, the term involving the double integral can be approximated
by $\int_{-\infty}^{\infty}dv\mathcal{R}\left(v\right)\int^{v}dv'\mathcal{R}\left(v'\right)=1/2\left.\left[\int^{v}dv'\mathcal{R}\left(v'\right)\right]^{2}\right|_{-\infty}^{\infty}=1/2$.
The growth rate, therefore, can be written as
\begin{equation}
\gamma_{L}\left(t\right)\simeq\gamma_{L0}\left\{ 1-\frac{\left|\omega_{b}^{2}\left(t\right)\right|^{2}}{2\nu^{4}}\left[\Gamma\left(\frac{1}{3}\right)\left(\frac{3}{2}\right)^{1/3}\frac{1}{3}-\frac{\pi}{2}\frac{\alpha^{2}}{\nu^{2}}\right]\right\} \label{eq:GammaL-1}
\end{equation}
which is the same expression calculated directly from nonlinear theory
(Ref. \citep{LestzDuartePoP2021}).

\section{Integration of the kinetic equation}

Since Eq. (2) of the Letter implies $f_{2}^{(0)}=0$, the kinetic
equation at first order in the amplitude parameter to be integrated
is 
\begin{multline}
\frac{\partial f_{1}^{(1)}}{\partial t}+i\left(kv-\omega\right)f_{1}^{(1)}+\frac{\omega_{b}^{2}F'_{0}}{2k}=\frac{\nu^{3}}{k^{2}}f_{1}''^{(1)}+\frac{\alpha^{2}}{k}f_{1}'^{(1)}\label{eq:KineticEqForN-1}
\end{multline}
Eq. \eqref{eq:KineticEqForN-1} has the solution 
\begin{multline}
f_{1}^{(1)}=-\frac{F'_{0}}{2k}\int_{0}^{t}dt_{1}\omega_{b}^{2}\left(t_{1}\right)\times\\
\times e^{-i\left(kv-\omega\right)\left(t-t_{1}\right)-i\alpha^{2}\left(t-t_{1}\right)^{2}/2-\nu^{3}\left(t-t_{1}\right)^{3}/3},\label{eq:f1scatt-1-1}
\end{multline}
Defining $s\equiv\nu\left(t-t_{1}\right)$,

\begin{equation}
f_{1}^{(1)}=-\frac{F'_{0}}{2\nu k}\int_{0}^{\nu t}ds\omega_{b}^{2}\left(t-s/\nu\right)e^{-i\left(\frac{kv-\omega}{\nu}\right)s-i\frac{\alpha^{2}}{\nu^{2}}s^{2}/2-s^{3}/3}.\label{eq:f1scatt-1}
\end{equation}
Due to the cubic exponential, only small values of $s\lesssim2$ contribute
significantly to the integral. 
Physically, when stochastic collisions dominate the system's dynamics
over other processes, the system memory is poorly retained and the
dynamics instead become essentially local in time \citep{duarte2018analytical,LestzDuartePoP2021},
which occurs when $\nu\gg\gamma_{L,0}-\gamma_{d}$. Therefore, the
amplitude can be evaluated at that peak of the exponential kernel
and be moved out of the integral and the upper limit of integration
can be set to infinity,
\begin{equation}
f_{1}^{(1)}=-\frac{F'_{0}\omega_{b}^{2}\left(t\right)}{2\nu k}\int_{0}^{\infty}dse^{-i\left(\frac{kv-\omega}{\nu}\right)s-i\frac{\alpha^{2}}{\nu^{2}}s^{2}/2-s^{3}/3}.\label{eq:f1scatt-1-2}
\end{equation}
Eq. \eqref{eq:f1scatt-1-2} is also a solution of Eq. \eqref{eq:KineticEqForN-1}
when the time derivative is neglected from the outset. That can be
easily seen by substituting Eq. \eqref{eq:f1scatt-1-2} into Eq. \eqref{eq:KineticEqForN-1},
which leads to
\begin{multline}
\frac{F'_{0}\omega_{b}^{2}\left(t\right)}{2k}=-\frac{F'_{0}\omega_{b}^{2}\left(t\right)}{2k}\int_{0}^{\infty}ds\times\\
\times\left[-i\left(\frac{kv-\omega}{\nu}\right)-i\frac{\alpha^{2}}{\nu^{2}}s-s^{2}\right]e^{-i\left(\frac{kv-\omega}{\nu}\right)s-i\frac{\alpha^{2}}{\nu^{2}}s^{2}/2-s^{3}/3},\label{eq:Equality}
\end{multline}
Noticing that the integrand of Eq. \eqref{eq:Equality} can be re-written
as $\frac{d}{ds}e^{-i\left(\frac{kv-\omega}{\nu}\right)s-i\frac{\alpha^{2}}{\nu^{2}}s^{2}/2-s^{3}/3},$
it follows straighforwardly that the equality (Eq. \eqref{eq:Equality})
is satisfied and Eq. \eqref{eq:f1scatt-1-2} is a solution of Eq.
\eqref{eq:KineticEqForN-1} without the time derivative.

\section{Generalization to tokamak geometry}

Axisymmetric tokamaks admit the kinetic energy $\mathcal{E}$, the
canonical toroidal momentum $P_{\varphi}$ and the magnetic moment
$\mu$ as invariants of the unperturbed particle motion. In the presence
of low-frequency (compared to the cyclotron frequency) perturbations,
the magnetic moment remains approximately preserved, while neither
$\mathcal{E}$ nor $P_{\varphi}$are conserved independently. Instead,
a new invariant emerges, $\ensuremath{\mathcal{E}^{\prime}=\mathcal{E}+\omega P_{\varphi}/n,}$
as long as the ansatz $e^{-i\left(n\varphi\omega t\right)}$ holds
for the wave perturbation. The existence of two invariants implies
that the resulting resonant particle motion is, therefore, one dimensional
in the vicinity of an isolated resonance determined by the condition
$\Omega=\omega+n\omega_{\varphi}-p\omega_{\theta}=0$, where $\omega_{\theta}$
and $\omega_{\varphi}$ are the mean poloidal and toroidal transit
frequencies of the unperturbed orbit, $p$ is an integer and $\omega$
is the frequency of a mode with toroidal number $n$. The fact that
the dynamics near a single resonance are one-dimensional allows the
idealized bump-on-tail system to be mapped into more realistic geometries,
as shown in Ref. \citep{BerkPPR1997}. In particular, modeling the
resonant interaction of fast ions with \Alfvenic waves in toroidal
geometry can be made more convenient by employing the relevant relaxation
variable $\Omega\left(P_{\varphi},\mathcal{E},\mu\right)$, which
directly maps onto the bump-on-tail framework (by replacing $kv-\omega$
with $\Omega$).

\section{Calculation of $\alpha/\nu$ in realistic tokamak scenarios }

As discussed in the Letter, heuristic arguments describe the enhancement
of the effective collisional frequencies $\alpha$ and $\nu$ within
a narrow resonance, relative to the inverse slowing down time $\tau_{s}^{-1}$
and 90-degree pitch angle scattering frequency $\nu_{\perp},$ respectively.
Combination of those scaling expressions with well-known formulas
for the macroscopic collision frequencies together yield a simple
expression for $\alpha/\nu$ in a uniform plasma, appropriate for
the bump-on-tail problem, which gives the relation analyzed in the
letter. However, in realistic scenarios with nontrivial magnetic geometries
and plasma profiles, such an expression is an incomplete accounting
of the competing collisional effects within the resonance. In a nonuniform
plasma, such as a tokamak, \Alfvenic eigenstructures can be spatially
localized, and its resonances with fast ions are spread out across
multiple surfaces in the $\left(P_{\varphi},\mathcal{E},\mu\right)$
constant-of-motion space. In this case, it is necessary to perform
averages when calculating $\alpha$ and $\nu$ over both the bounce
orbit motion and over the phase-space resonance surfaces weighted
by their relative contribution to driving the mode. 

In a tokamak, these averages depend on details of the equilibrium,
eigenmode properties, and fast ion distribution, and thus must be
performed numerically with a presently labor-intensive workflow. In
Refs. \citep{DuarteAxivPRL} and \citep{DuarteNF2018Likelihood},
$\alpha$ and $\nu$ were evaluated in this way for several shots
from NSTX, DIII-D, TFTR and ITER
, yielding much more accurate values than could be obtained from the
simple parametric scaling relation. 
There, it was found that different eigenmodes can have substantially
different values of $\alpha$ and $\nu$, even within the same time
of a discharge. Turbulence enters the calculation via an enhanced
diffusivity, utilizing scaling laws arising from gyrokinetic simulations
\citep{ZhangLinChen2008PRL,LangFu2011}. Whereas the inclusion of
turbulence significantly decreases the value of $\alpha/\nu$ in conventional
tokamaks, to the point of suppressing non-steady behavior, it hardly
affects the calculation of $\alpha/\nu$ in NSTX, as shown in Fig.
3 of Ref. \citep{DuartePoP2017}, due to the fact that ion transport
is typically near neoclassical levels in spherical tokamaks \citep{KayeNF2007}.
Consequently, it can be concluded that configurations with reduced
turbulence, such as spherical tokamaks and negative triangularity
plasmas, are more susceptible to the influence of drag and more closely
adhere to the scaling expression for $\alpha/\nu$ derived without
turbulent effects. Quantitatively reliable calculations of this ratio
require case-by-case calculation, even for a given tokamak. One situation
where values obtained from the heuristic expression alone can be considered
more reliable is a linear device, such as LAPD \citep{ZhangHeidbrinkLAPD_PoP2099,Heidbrink_LAPD_2012,Gekelman_LAPD_2016},
where the simpler geometry leads to less complicated resonance surfaces
(in addition to better control over the level of turbulent transport),
such that careful averaging is not as essential as in a tokamak. Due
to much lower plasma temperatures, laboratory plasmas such as those
studied in LAPD are a promising candidate for experimentally probing
the influence of drag in a controlled setting, with $\alpha/\nu\gtrsim5$.
The shifting and splitting of the resonance function can in principle
be detected by laser induced florescence measurements of the perturbed
distribution in a set-up similar to that of Ref. \citep{SarfatySkiffPRL1998}
for the case of resonant ions or using whistler mode wave absorption
for resonant electrons \citep{SchroderSkiff2017}.

\newcommand{\abs}[1]{\left|#1\right|} \newcommand{\br}{\text{Re}[b]} \newcommand{\bi}{\text{Im}[b]}

\section{Nonlinear theory in the time-local regime}
\label{sec:paperOne}

When the rate of scattering is much larger than the growth rate of the mode ($\nu\gg\gamma$), the evolution of the bump-on-tail problem becomes local in time, for any value of $\alpha/\nu$. In general, the evolution of the complex mode amplitude is given by the paradigmatic Berk-Breizman cubic equation 
\begin{multline} \frac{dA(\tau)}{d\tau} = A(\tau) - \frac{1}{2}\int_0^{\tau/2}dz z^2 A(\tau-z) \int_0^{\tau-2z}dx \\ e^{-\hat{\nu}^3 z^2 (2z/3 + x) + i \hat{\alpha}^2 z(z + x)}A(\tau-z-x)A^*(\tau-2z-x)  . \label{eq:cubic} \end{multline}
For compactness, normalized variables were introduced, with $A(\tau) = (\abs{\omega_b^2(\tau)}/\gamma^2)\sqrt{1 - \gamma_d/\gamma_{L,0}}$, $\gamma = \gamma_{L,0} - \gamma_d$, $\tau = t\gamma$, $\hat{\nu} = \nu/\gamma$, and $\hat{\alpha} = \alpha/\gamma$. When $\hat{\nu} \gg 1$, the integrand in Eq. \eqref{eq:cubic} is vanishingly small except near $x = 0$ and $z \approx 1/\hat{\nu} \ll 1$. In this case, the time delays present in the arguments of the mode amplitude become negligible, such that the product $A(\tau-z)A(\tau-z-x)A^*(\tau-2z-x) \approx A(\tau)\abs{A(\tau)}^2$ can be pulled out of the integral. This approximation reduces the time-delayed integro-differential equation for the mode amplitude into an ordinary differential equation, dubbed the time-local cubic equation since it makes explicit the irrelevance of the system's history in determining its future evolution:
\begin{align}
\label{eq:tlc}
\frac{dA(\tau)}{d\tau} &= A(\tau) - b(\hat{\alpha},\hat{\nu}) A(\tau)\abs{A(\tau)}^2, \quad\text{where }\\  b(\hat{\alpha},\hat{\nu}) &= \frac{1}{2\hat{\nu}^4}\int_0^\infty \frac{e^{-2 u^3/3 + i \hat{\alpha}^2 u^2/\hat{\nu}^2}}{1 - i\hat{\alpha}^2 / (\hat{\nu}^2 u)}du . \label{eq:bfull} \end{align}
A complete analytic solution for the time-local cubic equation was derived in Ref. \citep{LestzDuartePoP2021}. Of relevance here is its asymptotic behavior in the limit of $\tau \rightarrow \infty$, which gives regular oscillations at a fixed amplitude 
\begin{align} 	
\lim_{\tau\rightarrow\infty} A(\tau) = A_\text{sat} e^{-i\delta\omega_\text{sat}\tau/\gamma} = \frac{e^{-i(\bi/\br)\tau}}{\sqrt{\br}}
\label{eq:asymptotic} 
\end{align}
Hence the collisional constant $b$ contains information both on the saturation amplitude of the wave and drag-induced frequency shift $\delta\omega_\text{sat}$ (not discussed in the Letter). $A_\text{sat} = 1/\sqrt{\br}$ gives the wave saturation amplitude for all values of $\alpha/\nu < 0.96$. An equivalent expression for $A_\text{sat}$ was first derived by Lilley \citep{Lilley2009PRL}. Its expansion when $\alpha\ll\nu$ is given
by Eq. 26 of Ref. \citep{LestzDuartePoP2021}, which matches the saturation of the quasilinear system (Eq. 12) derived in the Letter exactly. 
The quasilinear system in the Letter emerges from the nonlinear system with two key assumptions: 1) $\abs{\omega_b^2} \ll \nu^2$, used to expand the Vlasov equation in orders of the mode amplitude and 2) the ''time-local'' approximation, enabling the time delays to be ignored in deriving the perturbed distribution function. The constraints implied by these assumptions will be outlined here. 
Define a new function $I^{-2}(\alpha/\nu) = \hat{\nu}^4 \br$, which isolates the dependence of the saturation level on the ratio $\alpha/\nu$. Then in order for $\abs{\omega_b^2} \ll \nu^2$ to be valid through mode saturation, one must require that 
\begin{align}  1 - \gamma_d/\gamma_{L,0} \ll \frac{1}{I^2(\alpha/\nu)} .  \label{eq:gdgl} \end{align} 
Since $A_\text{sat}$ (and therefore, $I^{-2}$) is an increasing function of $\alpha/\nu$, this shows that larger values of $\alpha/\nu$ require the system to be closer to marginal stability for the expansion to hold at all times. As points of reference, Eq. \eqref{eq:gdgl} requires  $1 - \gamma_d/\gamma_{L,0} \ll 0.51$ in the absence of drag, $1 - \gamma_d/\gamma_{L,0} \ll 0.33$ for $\alpha/\nu = 0.5$, and $1 - \gamma_d/\gamma_{L,0} \ll 0.03$ for $\alpha/\nu = 0.9$. For any value of $\alpha/\nu < 0.96$, there exist sufficiently small values of $1 - \gamma_d/\gamma_{L,0} \ll$ to satisfy Eq. \eqref{eq:gdgl}. 
Regarding the time-local assumption, neglecting the time delays is justified when $A$ does not change much between $\tau$ and $\tau - \delta$, where $\delta$ is a general time delay. Quantitatively, one must require $\abs{A(\tau - \delta/\hat{\nu}) - A(\tau)}/\abs{A(\tau - \delta/\hat{\nu})} \ll 1$, which can be rearranged for the maximum relevant time delay as $\abs{A'(\tau)}/\abs{A(\tau)} \ll \hat{\nu}/3$. Writing $A(\tau) = \abs{A(\tau)}e^{i\phi(\tau)}$, this condition becomes $\sqrt{(\abs{A}'/\abs{A})^2 + (\phi')^2} \ll \hat{\nu}/3$. Rigorous arguments to bound $\abs{\abs{A}'}/\abs{A}$ and $\abs{\phi'}$ can be found in Appendix A of Ref. \citep{LestzDuartePoP2021}, which is beyond the scope of this Supplemental Material. In a nutshell, $\abs{\abs{A}'}/\abs{A} < 1$ because the fastest growth of the mode occurs during the initial linear phase, and $\abs{\phi'} < \bi/\br$ because the fastest variation of the phase occurs in the saturation phase. Combination of these bounds gives 
\begin{align} 3\sqrt{1 + \frac{\bi^2}{\br^2}} \ll \frac{\nu}{\gamma} .  \label{eq:nuhat} \end{align}
Since $\bi/\br$ is an increasing function of $\alpha/\nu$, Eq. \eqref{eq:nuhat} demonstrates that larger values of $\alpha/\nu$ require relatively larger $\nu/\gamma$ in order to justify the time-local assumption. With only scattering, the condition becomes $\hat{\nu} \gg 3$. When $\alpha/\nu \approx 0.6$, the two terms under the radical are equal, and Eq. \eqref{eq:nuhat} requires $\hat{\nu} \gg 4.2$. As $\alpha$ approaches $\nu$, the constraint becomes more restrictive, with $\alpha/\nu = 0.9$ requiring $\hat{\nu} \gg 22$.  

\section{Sensitivity to the Marginality Parameter $\gamma_{d}/\gamma_{L,0}$}

\begin{figure}
\includegraphics[width=1\columnwidth]{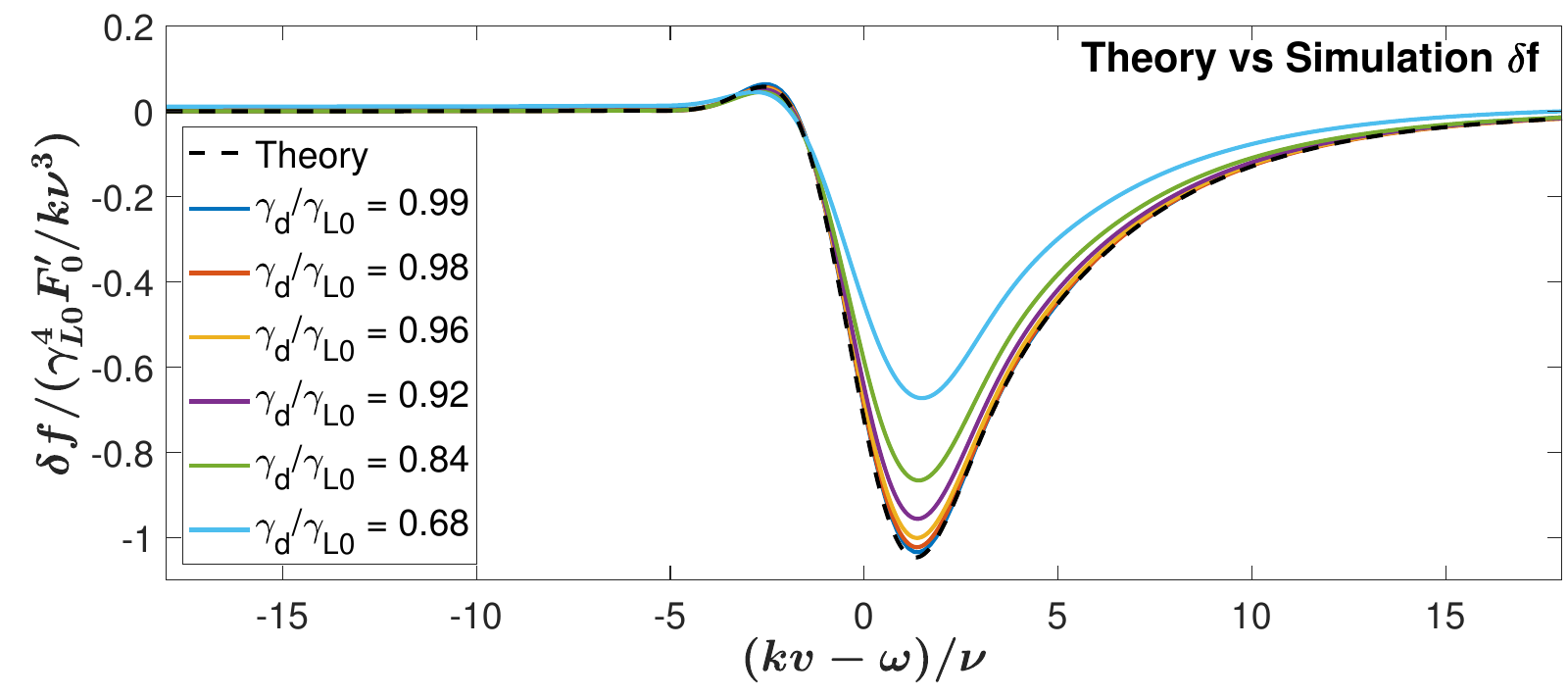}

\caption{Saturated distribution modification with respect to the initial equilibrium
distribution, $\delta f=f-F_{0}$. The colored curves are simulation
results from the nonlinear Vlasov code BOT using $\nu/(\gamma_{L,0}-\gamma_{d})$
= 20 and $\alpha/\nu=0.5$. The dashed curve is the theoretical result
for $1-\gamma_{d}/\gamma_{L,0}\ll1$. \label{fig:gdglscan}}
\end{figure}

In this section, we document a scan of the marginality parameter $1-\gamma_{d}/\gamma_{L,0}$,
which is assumed to be small in this work. This scan is shown in Fig.
\ref{fig:gdglscan}. There, simulations are performed using the BOT
code varying the marginality parameter while keeping the collisional
constants $\nu/(\gamma_{L,0}-\gamma_{d})=20$ and $\alpha/(\gamma_{L,0}-\gamma_{d})=10$
fixed. 
The theoretical prediction using a small parameter expansion in $1-\gamma_{d}/\gamma_{L,0}\ll1$
is overlayed as a dashed black curve. In this case, there is hardly
any discrepancy between theory and simulations until $\gamma_{d}/\gamma_{L,0}\leq0.84$.
Even then, the shape of $\delta f$ is still very similar, with its
magnitude requiring higher order corrections in the small parameter
for accuracy. Note that this value of $\gamma_{d}/\gamma_{L,0}\approx0.8$
is not a universal parameter where the theory is expected to break
down. As described in Sec. \ref{sec:paperOne}, the quantiative requirement
on $\gamma_{d}/\gamma_{L,0}$ for the theory to remain valid is itself
a function of $\alpha/\nu$ and $\nu/\gamma$. Specifically, for $\nu/\gamma=20$
and $\alpha/\nu=0.5$, Eqs. (\ref{eq:gdgl}) and (\ref{eq:nuhat})
require $1-\gamma_{d}/\gamma_{L,0}\ll0.33$ and $\nu/\gamma\gg3.7$,
respectively, for the theory to be valid. Hence it is consistent that
the simulations and theory begin to disagree when $1-\gamma_{d}/\gamma_{L,0}\gtrsim0.2$
for these parameters. 

\section{Sensitivity to the Collisionality Parameter $\nu/\gamma$}

\begin{figure}
\includegraphics[width=1\columnwidth]{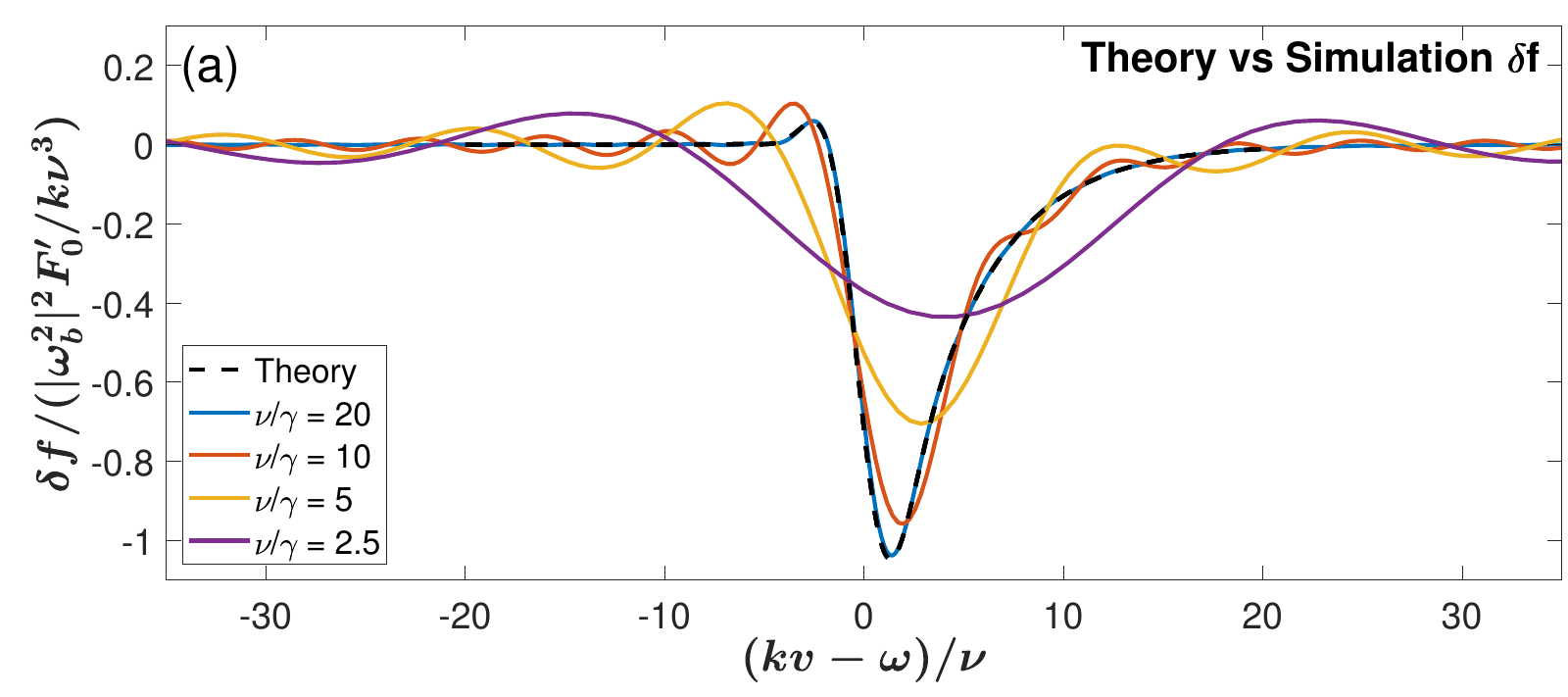}

\includegraphics[width=1\columnwidth]{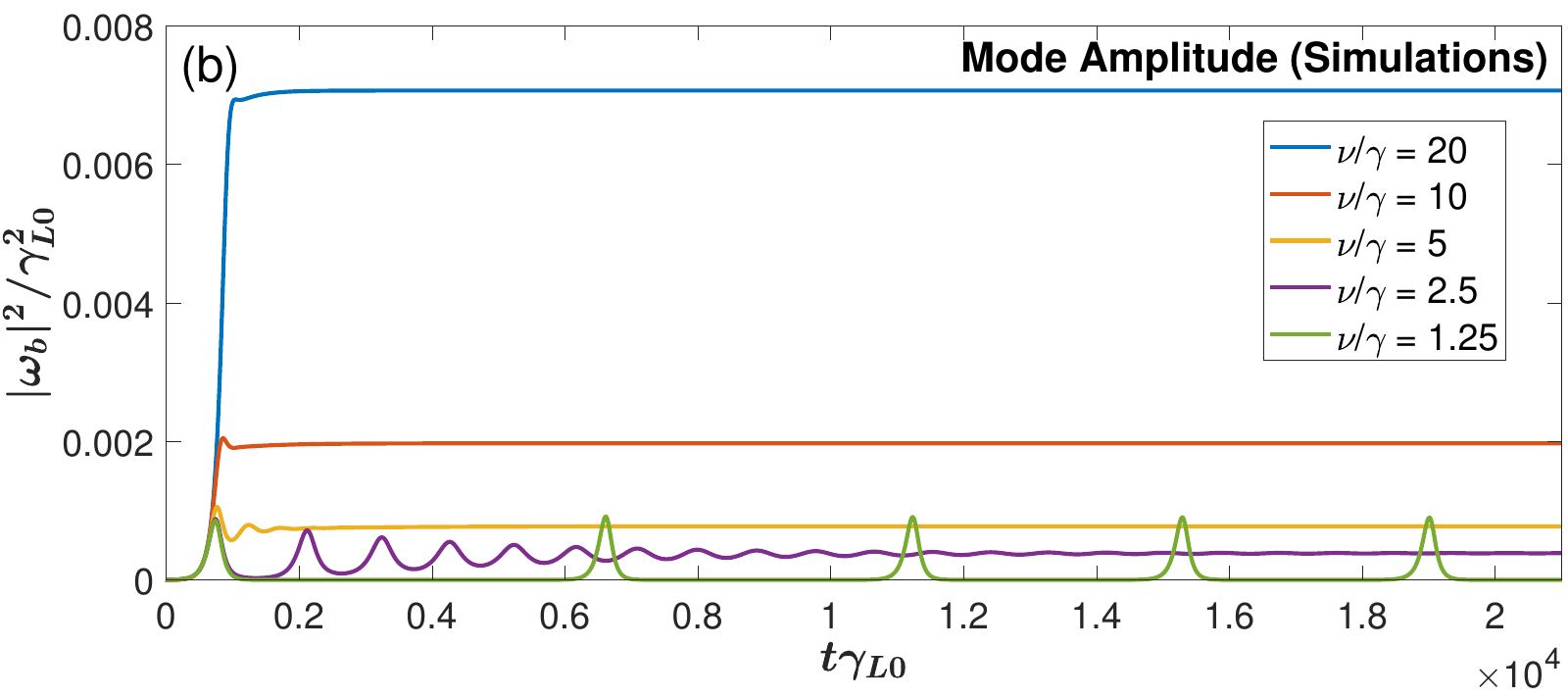}

\caption{(a) Saturated distribution modification with respect to the initial
equilibrium distribution, $\delta f=f-F_{0}$ and (b) mode amplitude
evolution as a function of time. The colored curves are simulation
results from the nonlinear Vlasov code BOT using $\alpha/\nu=0.5$
and $\gamma_{d}/\gamma_{L,0}=0.99$. The dashed curve in (a) is the
theoretical result for $\nu/\gamma\gg1$. \label{fig:nuscan}}
\end{figure}

In this section, we document a scan of the collisionality parameter
$\nu/\gamma$, which is assumed to be large in this work. BOT simulations
are performed with fixed values of $\gamma_{d}/\gamma_{L,0}=0.99$
and $\alpha/\nu=0.5$, while $\nu/\gamma$ is varied from 1.25 to
20. The perturbed distribution $\delta f$ at saturation in the simulations
is shown in Fig. \ref{fig:nuscan}(a) for $\nu/\gamma=2.5-20$. The
analytic expression for $\delta f$ in the limit of $\nu/\gamma\gg1$
is overlayed as the dashed black curve. Note that the plotted quantity
$\delta f$ is normalized by the squared mode amplitude (proportional
to $\left|\omega_{b}^{2}\right|^{2}\propto\nu^{4}$) in order to facilitate
comparison on the same scale. Good quantitative agreement is found
for $\nu/\gamma\geq10$, with differences appearing for $\nu/\gamma\leq5.$
This is consistent with the theoretical formalism, which involves
the implicit quantitative assumption of $\nu/\gamma\gg3.7$ for these
parameters (this threshold value depends on $\alpha/\nu$, as prescribed
by Eq. (\ref{eq:nuhat})). 

The corresponding time evolution of the mode amplitude is shown in
Fig. \ref{fig:nuscan}(b), extending down to $\nu/\gamma=1.25$. Consequently,
this parameter scan includes a transition from the collisional regime
where orbits are stochastic ($\left|\omega_{b}^{2}\right|\ll\nu^{2})$
to the strongly nonlinear regime where the resonant island structure
is preserved ($\left|\omega_{b}^{2}\right|\gg\nu^{2})$. To convert
from the plotted quantity to the normalized collisional parameter
$\left|\omega_{b}^{2}\right|/\nu^{2},$one would multiply by $\gamma^{2}/\nu^{2}(1-\gamma_{d}/\gamma_{L,0})^{2}$.
For sufficiently large values of $\nu/\gamma$, the linear growth
phase is followed almost immediately by saturation at a steady level,
with the collisional parameter remaining small ($\left|\omega_{b}^{2}\right|/\nu^{2}\approx0.18$
at saturation). As $\nu/\gamma$ is lowered, the saturation level
$\left|\omega_{b}^{2}\right|/\gamma_{L,0}^{2}$ decreases since$\left|\omega_{b,\text{{sat}}}\right|\propto\nu$
at constant $\alpha/\nu$, as in Eq. (12) of the Letter, and decaying
oscillations occur about the saturation level. At very small values
of $\nu/\gamma$, the initial growth phase transitions from the stochastic
to strongly nonlinear regime, and consequently the quasi-steady saturation
seen in simulations with larger $\nu/\gamma$ is replaced by pulsating
solutions. Since no steady state is achieved in these simulations
due to the pulsations, the $\delta f$ can not be compared in Fig.
\ref{fig:nuscan}(a) against the other simulations. To reiterate,
the focus of this work is on near threshold instabilities $(1-\gamma_{d}/\gamma_{L,0}\ll1)$
that are expected to be in the collisional regime since $\nu/\gamma$
is typically a large parameter when $\gamma$ is so small. However,
the results are also valid in the early phase of the instabilities
that will eventually enter the strongly nonlinear regime, up to the
point where $\left|\omega_{b}^{2}\right|\ll\nu^{2}$ is no longer
satisfied, which is where the near threshold ordering breaks down. 

\section{Conclusions and outlook}

This work has identified several ways in which the presence of collisional
drag fundamentally modifies the character of resonant wave-particle
interactions in kinetic systems and the associated transport. It has
been shown that near marginal stability and with sufficiently large
effective scattering, nonlinear theory naturally reduces to QL theory
even when coherence-preserving dynamical friction is included. This
finding provides further physical justification for the use of QL
models, such as the resonance broadened quasilinear code \citep{GorelenkovDuarteRBQNF2018},
for predicting fast ion transport due to \Alfven eigenmodes in otherwise
computationally intensive burning plasma scenarios. Of equal importance,
an isomorphism between the collisional bump-on-tail problem in plasmas,
studied in this work, and the resonant interaction of the rotating
galactic bar with orbit heavy bodies \citep{Hamilton2022} allows
these results and techniques to be applied to the distinct context
of galactic dynamics. The strength of the wave-particle interaction
is characterized by a collisionally-broadened resonance function.
The inclusion of drag breaks a symmetry otherwise present in the system,
shifting the resonance function from its usual location in phase space.
When drag dominates, the resonance function splits into multiple peaks
of similar magnitude. In particular, the particle redistribution is
found to be very sensitive to even small amounts of drag, indicating
the importance of including it in realistic models of resonant particle
transport. \bibliographystyle{apsrev4-1}
\addcontentsline{toc}{section}{\refname}